\documentclass[preprint,12pt,authoryear]{elsarticle}

\usepackage{url}
\usepackage{hyperref}
\usepackage{amsmath}
\usepackage{algorithm}
\usepackage{algpseudocode}
\journal{Nuclear Physics B}

\begin{document}

\begin{frontmatter}



\title{Accelerated Design of Mechanically Hard Magnetically Soft High-entropy Alloys via Multi-objective Bayesian Optimization} %% Article title


\author[tuda]{Mian Dai $^{\dagger}$} 
\author[tuda]{Yixuan Zhang $^{\dagger}$} 
\author[tuda]{Weijia He} 
\author[tuda]{Chen Shen\textsuperscript{*}}
\author[kth]{Xiaoqing Li} 
\author[kth]{Stephan Schönecker} 
\author[mpi]{Liuliu Han}
\author[tuda]{Ruiwen Xie\textsuperscript{*}}
\author[cup]{Tianhang Zhou\textsuperscript{*}}
\author[tuda]{Hongbin Zhang}

\affiliation[tuda]{organization={Institute of Materials Science, Technical University of Darmstadt},%Department and Organization
            % addressline={Alarich-Weiss-Str. 2}, 
            city={Darmstadt},
            country={Germany}}
\affiliation[kth]{organization={Department of Materials Science and Engineering, KTH - Royal Institute of Technology},%Department and Organization
            % addressline={SE-10044}, 
            city={Stockholm},
            country={Sweden}}
\affiliation[mpi]{organization={Max Planck Institute for Sustainable Materials},%Department and Organization
            % addressline={Max-Planck-Str. 1}, 
            city={Düsseldorf},
            country={Germany}}
\affiliation[cup]{organization={College of Carbon Neutrality Future Technology, China University of Petroleum},%Department and Organization
        % addressline={Max-Planck-Str. 1}, 
        city={Beijing},
        country={China}}                     
%% Abstract
\begin{abstract}
Designing high-entropy alloys (HEAs) that are both mechanically hard and possess soft magnetic properties is inherently challenging, as a trade-off is needed for mechanical and magnetic properties. In this study, we optimize HEA compositions using a multi-objective Bayesian optimization (MOBO) framework to achieve simultaneous optimal mechanical and magnetic properties. An ensemble surrogate model is constructed to enhance the accuracy of machine learning surrogate models, while an efficient sampling strategy combining Monte Carlo sampling and acquisition function is applied to explore the high-dimensional compositional space. The implemented MOBO strategy successfully identifies Pareto-optimal compositions with enhanced mechanical and magnetic properties. The ensemble model provides robust and reliable predictions, and the sampling approach reduces the likelihood of entrapment in local optima. Our findings highlight specific elemental combinations that meet the dual design objectives, offering guidance for the synthesis of next-generation HEAs.

\end{abstract}



%% Keywords

\end{frontmatter}



\section{Introduction}\label{sec1}

High-entropy alloys (HEAs), first introduced in 2004~\cite{yeh2004NanostructuredHighEntropy,cantor2004MicrostructuralDevelopment}, have emerged as a novel class of materials with exceptional mechanical and functional properties. These characteristics arise from their high configurational entropy and complex atomic interactions~\cite{cantor2014MulticomponentHigh,yeh2016OverviewHighEntropy,hsuClarifyingFourCore2024}. Inherent in the multi-functional nature of HEAs~\cite{miracle2017CriticalReviewa,george2019HighentropyAlloysa,han2024MultifunctionalHighentropy}, their magnetic properties have been explored recently~\cite{han2021UltrastrongDuctile} and HEAs are demonstrated to be promising mechanically hard and magnetically soft alloys~\cite{han2022MechanicallyStronga,han2024TwogigapascalstrongDuctile}. This makes them attractive alternatives to conventional magnetic alloys, which are often limited by brittleness and insufficient mechanical strength~\cite{chen2020NovelUltrafinegrained}. Achieving high-performance soft magnetic HEAs~\cite{pathak2025AdditiveManufacturing}, however, requires a careful balance among magnetization, coercivity, ductility, and thermal stability. The vast compositional space combined with intricate property interdependencies~\cite{kormann2015TreasureMaps,lucas2011MagneticVibrationala} presents a formidable challenge for systematic discovery and optimization. Conventional trial-and-error approaches are neither scalable nor sufficiently systematic to explore such a high-dimensional landscape, underscoring the need for data-driven design strategies tailored to the complexity of HEAs.

First-principles methods such as density functional theory (DFT)\cite{tian2013InitioInvestigation,huang2018TwinningMetastable,zhang2018ElasticProperties}, in combination with thermodynamic methods\cite{senkov2015AcceleratedExploration,gorsse2018ReliabilityCALPHAD,zhang2016CALPHADModeling,shen2025synergy}, have established a rigorous foundation for understanding the structural and energetic behavior of HEAs. Building on this foundation, high-throughput (HTP) computational workflows~\cite{lederer2018SearchHigh,feng2021HighthroughputDesign,fukushima2022AutomaticExhaustive,chen2023MapSinglephase,dai2025DataDrivenDesign,shen2025multifunctionality,singh2021multifunctional,shen2021designing} have accelerated compositional screening and generated large datasets that facilitate the identification of promising candidates. Nevertheless, such exhaustive approaches remain resource-intensive and inefficient, particularly in high-dimensional spaces where significant effort may be wasted on suboptimal regions. To overcome these limitations, machine learning (ML) techniques~\cite{pei2020MachinelearningInformeda,rickman2020MachineLearninga,qiao2021FocusedReviewa,liu2023MachineLearningb,raabe2023AcceleratingDesigna} have been increasingly integrated into HTP pipelines, offering data-efficient strategies for materials discovery. By leveraging existing datasets, ML models have successfully accelerated single-objective optimization, enabling rapid identification of high-performing alloys~\cite{wen2019MachineLearninga,yang2022MachineLearningbased,liu2022AcceleratedDevelopment,giles2022MachinelearningbasedIntelligent}. 

To further enhance the efficiency of data-driven discovery, active learning strategies~\cite{xue2016AcceleratedSearcha,balachandran2018ExperimentalSearch,gubernatis2018MachineLearning,yuan2018AcceleratedDiscovery,lookman2019ActiveLearninga} have emerged as powerful tools for guiding the efficient navigation of HEA design spaces. In particular, Bayesian optimization (BO)\cite{snoek2012PracticalBayesian,frazier2016BayesianOptimization,shahriari2016TakingHumana,shen2025supersalt} has proven particularly effective for objectives that are costly to evaluate and limited in data availability. BO operates by iteratively constructing a surrogate model, typically a Gaussian process (GP)\cite{rasmussen2005GaussianProcesses}, to approximate the objective function and quantify uncertainty. An acquisition function~\cite{frazier2018TutorialBayesianb} then determines the next candidate by trading off exploration of uncertain regions against exploitation of promising areas. As new data become available, both the surrogate and the acquisition strategy are updated, steering the search toward optimal materials~\cite{ramprasad2017MachineLearning,talapatra2018AutonomousEfficient,ghoreishi2019EfficientUse,khatamsaz2021EfficientlyExploiting}. Recent studies have successfully applied BO to accelerate the discovery of functional HEA systems, including Invar-type alloys~\cite{rao2022MachineLearning}, metallic glasses~\cite{shi2023ConnectingCompositiona}, and electrocatalysts for the oxygen reduction reaction (ORR)\cite{pedersen2021BayesianOptimizationa}. The extension of BO to multi-objective frameworks has further enabled the simultaneous optimization of multiple, often competing, material properties through the construction of the Pareto front (PF)\cite{kim2023ExploringOptimal}. Notable examples include the design of ductile refractory HEAs under processing and physical constraints~\cite{khatamsaz2022MultiobjectiveMaterials,khatamsaz2023BayesianOptimization}, as well as the integration of graph neural networks with DFT data for identifying high-performance HEA catalysts~\cite{xu2024DiscoveringHigh}. Despite these advances, conventional BO approaches often encounter mode collapse in high-dimensional design spaces, which restricts their ability to explore diverse optimal solutions. Likewise, forward-model-based identification strategies~\cite{butler2018MachineLearning} are intrinsically tied to the limitations of the underlying dataset, frequently converging to locally optimal descriptions that hinder the discovery of superior candidates beyond the observed distribution.

We propose a scalable multi-objective Bayesian optimization (MOBO) framework, focusing on the mechanical and magnetic properties of HEAs. Instead of relying on traditional GP surrogates, our approach employs an ensemble-based model that naturally accommodates high-dimensional inputs with improved training and inference efficiency. To overcome the restrictions of fixed-element combinations, we introduce a Monte Carlo sampling strategy within a 10-element chemical space, allowing arbitrary subsets of five elements to be explored. This expands the search space and promotes the discovery of more diverse alloys. In addition, we relax the reliance on closed-form acquisition functions by numerically approximating standard metrics such as expected improvement and hypervolume improvement through repeated sampling from the ensemble surrogate. This sampling-based formulation supports flexible estimation of PF distributions and enables uncertainty-aware decision-making even when analytical tractability is lacking~\cite{balandat2020BoTorchFramework}.

We demonstrate the framework by optimizing Pugh's ratio, Cauchy pressure, saturation magnetization, and Curie temperature within quinary HEA systems from a pool of ten elements. Our approach shows its effectiveness in navigating complex design spaces under multi-objective constraints. These results highlight ensemble-based Bayesian optimization as a promising paradigm for next-generation materials discovery.

\section{Results}

\begin{figure}[h!]
\centering
\includegraphics[width=1\textwidth]{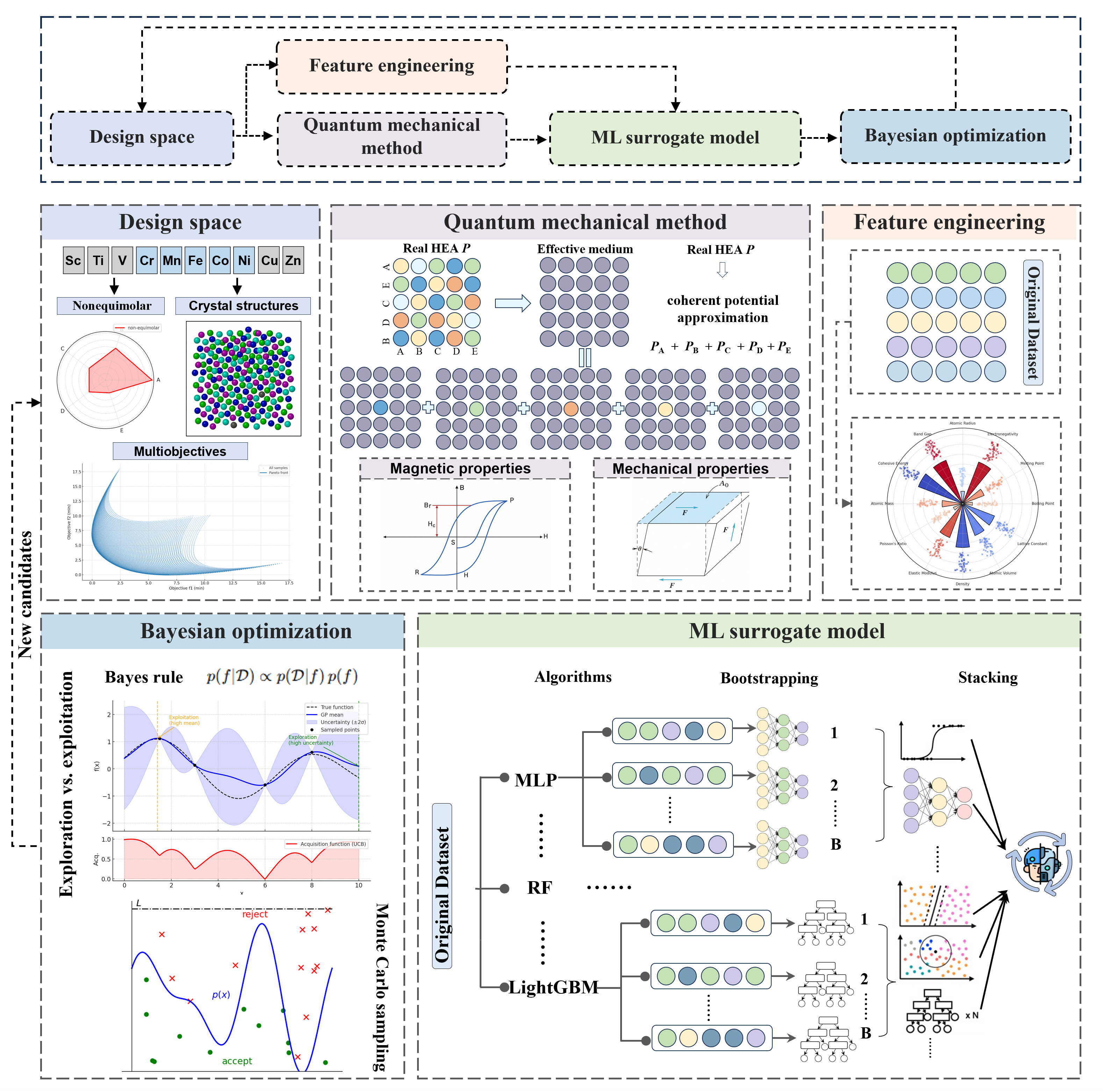}
\caption{Schematic overview of the multi-objective Bayesian optimization (MOBO) framework for high-entropy alloy (HEA) design. The workflow integrates four key components: (i) Design space, where candidate alloys are generated from a 10-element chemical pool with nonequimolar compositions and different crystal structures; (ii) Quantum mechanical method, where the coherent potential approximation treats chemical disorder and enables evaluation of magnetic and mechanical properties; (iii) Feature engineering, where datasets from quantum calculations are transformed into descriptors for machine learning; (iv) Machine learning (ML) surrogate model (e.g., multilayer perceptrons (MLPs), random forests (RFs), and gradient-boosted decision trees (LightGBM)), where ensemble learning with bootstrapping and stacking provides predictive accuracy and uncertainty quantification. These ML models are embedded into the MOBO loop, balancing exploration and exploitation to iteratively suggest new HEA candidates until convergence.}
\label{fig1}
\end{figure}

The framework implements a closed-loop workflow that integrates design space exploration, first-principles calculations, feature engineering, surrogate modeling, and Bayesian optimization in a sequential yet iterative manner as illustrated in figure~\ref{fig1}. The design space consists of a ten-element chemical pool ({Sc, Ti, V, Cr, Mn, Fe, Co, Ni, Cu, Zn}), from which not only equimolar but also non-equimolar five-element compositions are generated. Each candidate alloy is assigned to specific prototype crystal structures (e.g., body-centered cubic (BCC), face-centered cubic (FCC)) and multiple target objectives (magnetic moment, Curie temperature, Pugh’s ratio, and Cauchy pressure), resulting in a high-dimensional optimization problem under multiple conflicting objectives.

To investigate the composition–property relationships, first-principles calculations are carried out within DFT using the generalized gradient approximation (GGA) in combination with the coherent potential approximation (CPA). The CPA constructs a self-consistent effective medium that captures substitutional disorder in multi-component alloys. Based on this description, magnetic properties such as the ground-state magnetization and the Curie temperature evaluated within the disordered local moment (DLM) approach~\cite{gyorffy1985FirstprinciplesTheory}, as well as mechanical properties derived from the elastic constants, including Pugh’s ratio and Cauchy pressure, are systematically computed.

The processed datasets are subjected to feature engineering, in which compositional and structural descriptors are systematically transformed into statistically derived features (such as means and variances) and physically motivated quantities (including atomic radius, electronegativity, and valency)~\cite{wardGeneralpurposeMachineLearning2016}. These features serve as structured inputs for surrogate models and provide both statistical robustness and physical interpretability. A heterogeneous ensemble of base learners is constructed by combining multiple algorithmic families, such as decision trees, gradient boosting methods, and neural networks. Each base model is trained on bootstrap-resampled subsets of the data to enhance stability. The predictions from individual learners are subsequently aggregated through a stacking strategy, in which a meta-model integrates their outputs to achieve calibrated predictions and provide reliable uncertainty quantification~\cite{lakshminarayanan2017SimpleScalable}.

The Bayesian optimization module employs the ensemble surrogate’s posterior predictive distribution to select new alloy candidates in batches at each iteration. Candidate compositions are sampled in the 10-element simplex using a Monte Carlo scheme. For each candidate in the batch, the acquisition value is computed by maximizing the expected hypervolume improvement (EHVI). The newly acquired property data are incorporated into the training set, and the ensemble surrogate is retrained after every iteration to refine predictive accuracy. The iterative loop is terminated when the hypervolume gain falls below 0.1\% for ten consecutive iterations. In practice, no appreciable gain was observed beyond iteration 15, implying that the search would have been automatically terminated according to this criterion. Upon completion, the non-dominated set of alloys identified during the search and the corresponding estimated PF are reported.

\begin{figure}[h!]
\centering
\includegraphics[width=0.8\textwidth]{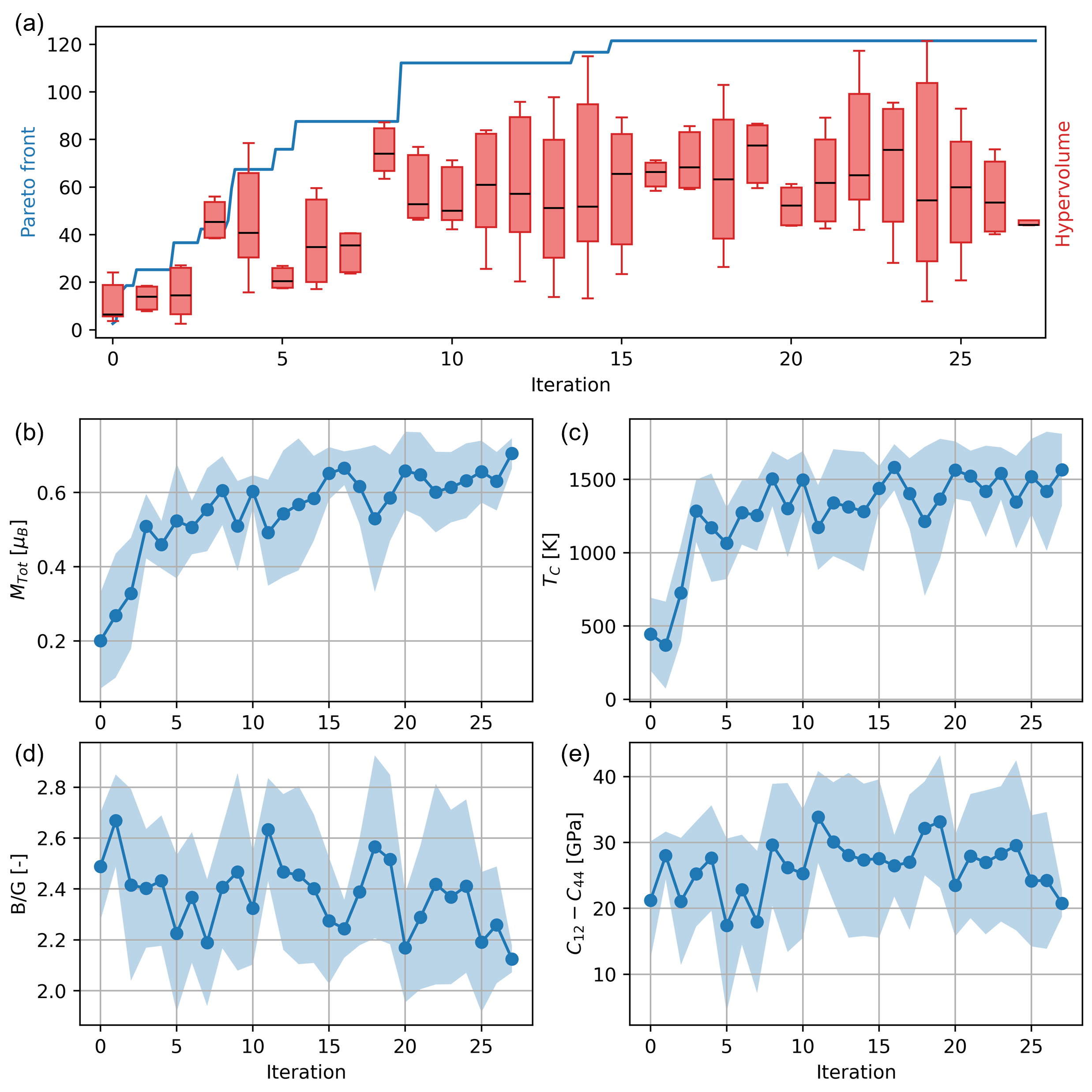}
\caption{
Convergence of the multi-objective search and per-property learning dynamics. (a) Best-so-far Pareto-front indicator (blue, left axis) increases monotonically with iteration, while red box-and-whisker plots report the distribution of attained hypervolume (right axis) across evaluations in each iteration, evidencing a rapid expansion of the dominated objective space followed by saturation. (b–e) Iteration-wise trajectories of the four target properties: total magnetic moment $M_{\mathrm{Tot}} [\mu_B]$, Curie temperature $T_C$ [K], Pugh's ratio $B/G$ and Cauchy pressure $C_{12}$ - $C_{44}$ [GPa]. Dots denote the per-iteration mean over evaluated candidates; shaded envelopes indicate the confidence interval computed from the samples in each iteration.
}
\label{fig2}
\end{figure}

Figure~\ref{fig2} evaluates the optimization process by tracking the evolution of the estimated Pareto front and the variability of hypervolume across BO iterations. Panel~(a) shows the best-so-far Pareto front, which rises in a stepwise monotonic pattern, indicating continual improvement of the non-dominated set. In parallel, the box plots report the distribution of hypervolume values within each iteration, where each batch consists of \(n=10\) evaluated candidates. In the first five iterations, hypervolume shows large jumps, whereas later iterations display smaller gains and finally approach a plateau. This trend reflects convergence toward a stable Pareto set under the given surrogate model and acquisition rule. Although the average hypervolume stabilizes, the per-iteration spread remains wide, as shown by the length of the box plots. This spread indicates that the acquisition strategy preserves diversity within each batch.

Panels~(b–e) show the evolution of batch means and confidence intervals for the four objectives. These objectives are the total magnetic moment \(M_{\mathrm{Tot}}\), the Curie temperature \(T_{\mathrm{C}}\,[\mathrm{K}]\), Pugh's ratio \(B/G\), and the Cauchy pressure \(C_{12}-C_{44}\,[\mathrm{GPa}]\). Each value is computed over a batch of ten compositions. The magnetic objectives in panels~(b) and (c) rise toward their target ranges, and their confidence intervals narrow steadily. This pattern indicates that the surrogate model rapidly learns the dominant composition–property relations. In contrast, the mechanical objectives in panels~(d) and (e) follow non-monotonic paths and retain wider confidence intervals across most iterations. This behavior is consistent with a rougher response surface and stronger trade-offs against the magnetic objectives. At several points, the confidence intervals widen abruptly, which coincides with exploratory batches in sparsely sampled regions. These episodes illustrate how the ensemble-based active learning strategy balances exploitation in well-understood regions with exploration where uncertainty is high as the search approaches Pareto stability.

\begin{table}[h!]
\centering
\caption{The set of the top 20 candidates that lie on the four-objective magnetic and mechanical Pareto front identified by the proposed framework.}
\label{tab:top20}
\footnotesize
\begin{tabular}{c|c|c|c|c}
 \hline
 Formula &  $M_{\mathrm{Tot}}\, [\mu_B/\mathrm{site}]$ &  $T_C\,[\mathrm{K}]$ &  $B/G$ &  $C_{12}-C_{44}\,[\mathrm{GPa}]$ \\
 \hline
Mn$_{0.064}$Fe$_{0.181}$Co$_{0.552}$Ni$_{0.091}$Cu$_{0.112}$ &  0.628 & 1449.591 & 2.790 & 44.564 \\
Mn$_{0.019}$Fe$_{0.396}$Co$_{0.320}$Ni$_{0.163}$Cu$_{0.102}$ &  0.653 & 1433.499 & 2.693 & 45.377 \\
Mn$_{0.058}$Fe$_{0.170}$Co$_{0.635}$Cu$_{0.117}$Zn$_{0.020}$ &  0.640 & 1690.254 & 2.597 & 40.579 \\
Mn$_{0.022}$Fe$_{0.309}$Co$_{0.365}$Ni$_{0.049}$Cu$_{0.255}$ &  0.568 & 1523.536 & 2.787 & 42.579 \\
Mn$_{0.035}$Fe$_{0.275}$Co$_{0.373}$Ni$_{0.086}$Cu$_{0.231}$ &  0.562 & 1400.610 & 2.866 & 43.266 \\
Ti$_{0.002}$Fe$_{0.491}$Co$_{0.198}$Ni$_{0.112}$Cu$_{0.197}$ &  0.635 & 1444.628 & 2.671 & 43.542 \\
Sc$_{0.034}$Fe$_{0.240}$Co$_{0.569}$Ni$_{0.078}$Cu$_{0.079}$ &  0.618 & 1672.585 & 2.601 & 39.838 \\
Mn$_{0.041}$Fe$_{0.269}$Co$_{0.438}$Cu$_{0.236}$Zn$_{0.016}$ &  0.585 & 1591.886 & 2.680 & 40.678 \\
Mn$_{0.039}$Fe$_{0.321}$Co$_{0.522}$Ni$_{0.042}$Cu$_{0.076}$ &  0.711 & 1737.375 & 2.413 & 39.645 \\
Sc$_{0.019}$Fe$_{0.472}$Co$_{0.217}$Ni$_{0.183}$Cu$_{0.109}$ &  0.651 & 1394.089 & 2.623 & 43.315 \\
V$_{0.023}$Fe$_{0.332}$Co$_{0.445}$Ni$_{0.108}$Cu$_{0.092}$  &  0.627 & 1561.996 & 2.576 & 40.056 \\
Mn$_{0.047}$Fe$_{0.337}$Co$_{0.212}$Ni$_{0.149}$Cu$_{0.255}$ &  0.546 & 1160.344 & 2.972 & 44.687 \\
Ti$_{0.021}$Fe$_{0.388}$Co$_{0.237}$Ni$_{0.207}$Cu$_{0.147}$ &  0.567 & 1258.288 & 2.828 & 43.933 \\
Mn$_{0.050}$Fe$_{0.279}$Co$_{0.312}$Ni$_{0.072}$Cu$_{0.287}$ &  0.540 & 1280.241 & 2.899 & 42.106 \\
Mn$_{0.019}$Fe$_{0.229}$Co$_{0.533}$Cu$_{0.168}$Zn$_{0.051}$ &  0.573 & 1706.580 & 2.608 & 36.254 \\
Ti$_{0.024}$Fe$_{0.371}$Co$_{0.341}$Ni$_{0.148}$Cu$_{0.116}$ &  0.611 & 1454.804 & 2.613 & 40.031 \\
Mn$_{0.031}$Fe$_{0.406}$Co$_{0.339}$Ni$_{0.032}$Cu$_{0.192}$ &  0.650 & 1587.271 & 2.507 & 38.032 \\
Mn$_{0.069}$Fe$_{0.239}$Co$_{0.512}$Cu$_{0.145}$Zn$_{0.035}$ &  0.631 & 1630.616 & 2.505 & 37.373 \\
Mn$_{0.050}$Fe$_{0.368}$Co$_{0.238}$Ni$_{0.007}$Cu$_{0.337}$ &  0.559 & 1375.574 & 2.763 & 38.901 \\
Mn$_{0.054}$Fe$_{0.332}$Co$_{0.227}$Ni$_{0.129}$Cu$_{0.258}$ &  0.559 & 1178.559 & 2.873 & 41.472 \\
\hline
\end{tabular}
\end{table}

The optimization produces a set of twenty high-performing candidates located on the four-objective Pareto front. Table~\ref{tab:top20} reports these alloys, selected by their top hypervolume contributions. All candidates stabilize in the BCC structure. The highest-ranked alloys are predominantly Fe–Co rich, often containing Mn and Ni, and they achieve the largest values of \(M_{\mathrm{Tot}}\) and \(T_C\). In contrast, alloys that score strongly on the mechanical properties \(B/G\) and \(C_{12}-C_{44}\) often include significant Cu fractions, whereas Zn and Ti occur only in small amounts among the top candidates. The most balanced trade-offs are concentrated within the Co–Fe–Mn–Ni–Cu space, with occasional minor contributions from Sc, Ti, or V. Across these twenty candidates, \(M_{\mathrm{Tot}}\) ranges from \(0.54\) to \(0.71\,\mu B\) , \(T_C\) from \(1{,}160\) to \(1{,}737\,\mathrm{K}\) with several exceeding \(1{,}600\,\mathrm{K}\), \(B/G\) from \(2.41\) to \(2.97\), and \(C_{12}-C_{44}\) from \(36\) to \(45\,\mathrm{GPa}\). These candidates represent promising targets for experimental validation of their combined soft-magnetic performance and mechanical strength. First-principles calculations show that adding Cu to Fe-Mn-(Cr, Co, Ni, Cu) alloys can raise the \(B/G\) ratio by nearly 19\% and enhance the magnetic moments of Fe and Mn~\cite{chen2025InfluenceFe}, while further investigations on FeCoNiCu-based high-entropy alloys reveal that alloying with Mn, V, and Cr significantly modulates both elastic and magnetic properties~\cite{huang2017ThermalExpansiona}. FeNiCoMn-based HEAs display higher saturation magnetization and Curie temperatures as Cu content increases~\cite{rao2020UnveilingMechanism}, similar to our findings of strong magnetization in Co-Fe-Mn-Ni-Cu alloys.

\begin{figure}[h!]
\centering
\includegraphics[width=0.8\textwidth]{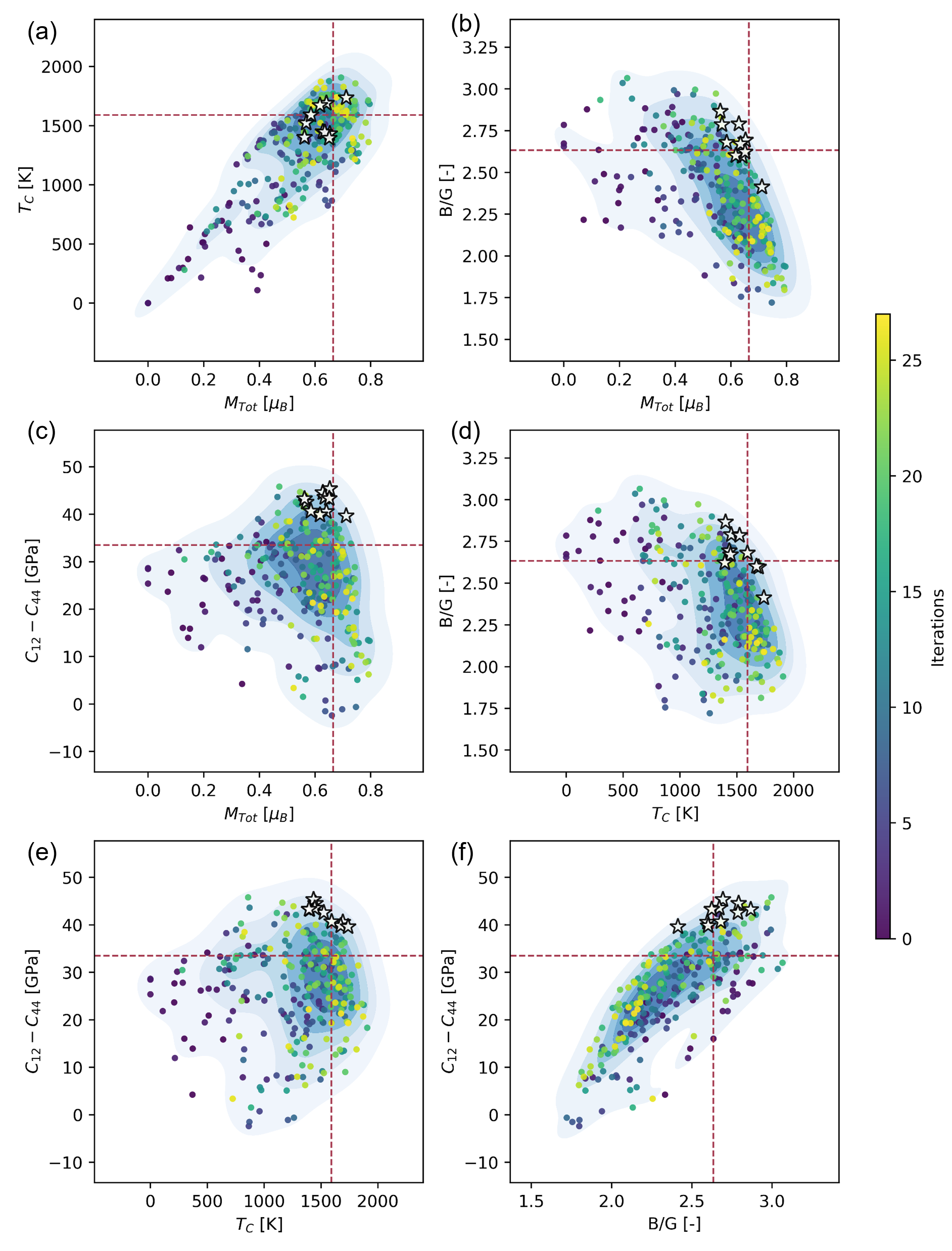}
\caption{
Pairwise objective landscapes obtained from multi-objective Bayesian optimization of high-entropy alloys. 
Each panel (a–f) shows the joint distribution of candidate alloys across two objectives: 
(a) Curie temperature $T_{\mathrm{C}}$ vs.\ total magnetic moment $M_{\mathrm{Tot}}$, 
(b) Pugh’s ratio $B/G$ vs.\ $M_{\mathrm{Tot}}$, 
(c) Cauchy pressure $C_{12}-C_{44}$ vs.\ $M_{\mathrm{Tot}}$, 
(d) $B/G$ vs.\ $T_{\mathrm{C}}$, 
(e) $C_{12}-C_{44}$ vs.\ $T_{\mathrm{C}}$, 
and (f) $C_{12}-C_{44}$ vs.\ $B/G$. 
Blue dots denote sampled candidates, shaded contours indicate density levels, and black star markers highlight Pareto-optimal solutions. 
Red dashed lines represent reference thresholds used to delineate desirable property regimes.
}
\label{fig3}
\end{figure}

To analyze the distribution of candidate solutions in the multi-objective design space, figure~\ref{fig3} presents pairwise projections of the four optimization objectives. Each panel shows the relationship between two objectives, making potential trade-offs and synergies visible across different property dimensions. The shaded density contours indicate the statistical distribution of evaluated compositions, and the markers represent the sampled candidates. Star symbols mark Pareto-optimal solutions, which consistently appear within confined subregions of each projection. The clustering of Pareto-optimal candidates across multiple property pairs shows that high-performing alloys concentrate in well-defined domains rather than being scattered randomly.

The consistent localization of Pareto-optimal candidates highlights two key features of the Bayesian optimization framework. First, the method directs the search away from unpromising regions of the objective space and toward areas where multiple properties can be improved simultaneously. Second, the recurrent overlap of Pareto-optimal candidates across different pairwise views demonstrates the coherence of the optimization results. This pattern suggests that the algorithm successfully identifies a subset of compositions that jointly optimize magnetic and mechanical properties. Overall, figure~\ref{fig3} shows that the Bayesian optimization strategy not only discovers individual high-performing candidates but also defines a robust region of the design space where optimal trade-offs and targeted convergence are achieved.

\begin{figure}[h!]
\centering
\includegraphics[width=0.8\textwidth]{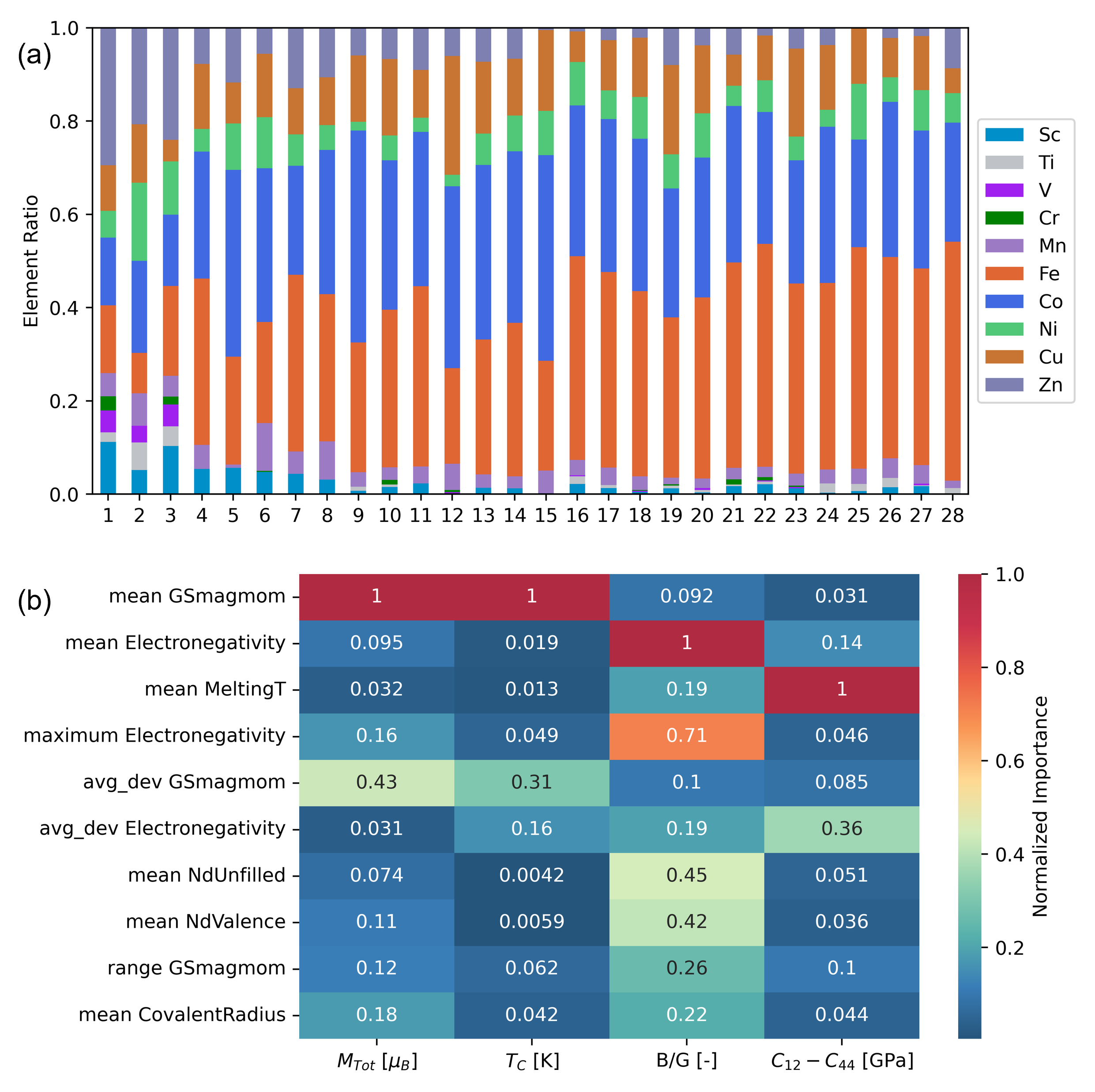}
\caption{
Evolution of composition and feature importance. (a) Averaged elemental compositions of candidate alloys over consecutive groups of ten optimization iterations, shown as stacked bar plots with color-coded contributions from the ten constituent elements (Sc, Ti, V, Cr, Mn, Fe, Co, Ni, Cu, Zn). Each bar represents the normalized mean element ratio within one batch of ten evaluated compositions. (b) Normalized feature importance of selected elemental descriptors with respect to the four optimization objectives: saturation magnetization $M_{s}$, Curie temperature $T_{C}$, Pugh’s ratio $B/G$, and Cauchy pressure $C_{12}-C_{44}$. 
}
\label{fig4}
\end{figure}

In order to better understand the origins of these Pareto-localized patterns, we next analyze how the optimization process shapes compositional preferences and descriptor-level drivers of performance. Figure~\ref{fig4}a tracks the evolution of batch-averaged elemental fractions across iterations to reveal preferential sampling in the compositional space. Elemental fractions are normalized to unity within each composition and averaged over batches of 10 candidates per iteration. As the search progresses, the proportions of Co, Fe, and Mn increase steadily, often exceeding 50\% of the batch composition in later iterations, while Zn, Ti, and V decline to below 5\% or fluctuate without a stable trend. To quantify this adaptive concentration, elemental frequencies are compared between early iterations (1–20) and later iterations ($>20$). In the later stage, Co- and Fe-containing systems dominate, together with consistent representation of Mn, Cu, and Ni, whereas Zn, Ti, and V occur only rarely. This adaptive shift indicates that the optimization loop systematically directs sampling toward subspaces more likely to deliver favorable trade-offs between magnetic and mechanical properties. 

Figure~\ref{fig4}b shows normalized importance scores of the selected descriptors for each target. Descriptors for mean ground-state magnetic moment and mean electronegativity rank highest for \(M_{\mathrm{Tot}}\) and \(T_C\). In contrast, descriptors such as the standard deviation of electronegativity and the variance of covalent radius are more important for \(B/G\) and \(C_{12}-C_{44}\).These associations are physically consistent, as average magnetic character and charge-transfer propensity support magnetic performance, whereas bonding heterogeneity and size mismatch act as proxies for ductility-related responses.

\section{Discussion}

The present study demonstrates that the proposed ensemble-based MOBO framework can efficiently navigate a ten-element HEA design space to simultaneously optimize magnetic and mechanical properties. Within fifteen iterations, the HV expands by more than 80\% of its final value and then stabilizes, indicating sample-efficient convergence under the current configuration. The trajectories of individual objectives reveal distinct learning dynamics: the magnetic targets (\(M_{\mathrm{Tot}}\) and \(T_{\mathrm{C}}\)) shift decisively toward desirable regimes with progressively narrowing confidence intervals, while the mechanical targets (\(B/G\) and \(C_{12}-C_{44}\)) follow non-monotonic paths and retain broader uncertainty bands. This contrast suggests that magnetic behavior is captured more readily than mechanical trade-offs, consistent with smoother response surfaces for the former and rougher landscapes for the latter. Pareto-optimal solutions cluster within a composition dominated by Co, Fe, Mn, Ni, and Cu, whereas Zn, Ti, and V are progressively de-emphasized, pointing to a physically plausible subspace for balanced design.

The enrichment of Co, Fe, and Ni correlating with higher \(M_{\mathrm{Tot}}\) and \(T_{\mathrm{C}}\) agrees with experimental studies of Co–Fe–Ni based soft magnetic HEAs \cite{han2022MechanicallyStronga}, indicating that the MOBO framework can reproduce known motifs within limited iterations. Likewise, the beneficial role of Cu in improving mechanical properties is consistent with CALPHAD~\cite{wen2024AchievingStrengthductility} and experimental~\cite{agrawal2022RoleCu,yuan2024CuAlloying} evidence that Cu additions enhance toughness in multi-component alloys.

The systematic exclusion of Zn, Ti, and V is also in line with conventional alloying strategies, since Zn tends to form brittle phases due to its low melting point~\cite{tsai2014HighEntropyAlloysa}, Ti and V often promote excessive lattice distortion and hard secondary phases that compromise ductility and soft-magnetic behavior~\cite{poletti2014ElectronicThermodynamic,george2020HighEntropy}, making these elements undesirable in the context of combined magnetic and mechanical performance. These comparisons underscore the ability of the framework not only to identify high-performing candidates but also to uncover results that align with established metallurgical intuition.

Beyond reproducing known systems, the present analysis reveals a new design insight. Dispersion-oriented descriptors, such as the standard deviation of electronegativity and the variance of covalent radius, strongly influence mechanical properties \cite{vazquez2022EfficientMachinelearning, ding2018TunableStacking, ma2024ChemicalInhomogeneities}. In contrast, mean-valued descriptors dominate magnetic performance. This suggests that controlling descriptor dispersion is a promising tool for engineering mechanically compatible soft magnetic HEAs. Moreover, the adaptive convergence of sampling toward the Co–Fe–Mn–Ni–Cu system demonstrates that the MOBO framework does not merely optimize numerically but also steers the search toward a physically meaningful subspace, reinforcing the interpretability of the approach.

Several limitations of the present study should be acknowledged. Critical quantities such as coercivity and magnetocrystalline anisotropy energy are not yet included, and microstructural or processing effects remain outside the current scope \cite{coey2010MagnetismMagnetic,gutfleisch2011MagneticMaterials}. The MOBO framework should be extended to incorporate such end-use metrics. 
The EMTO-CPA description combined with mean-field estimates of \(T_{\mathrm{C}}\) provides a tractable framework for disordered alloys but may overestimate Curie temperatures and neglect short-range order effects. While relative rankings are preserved, benchmarking against experimental reference systems is necessary to calibrate error bars \cite{kormann2015TreasureMaps}. Experimental synthesis and characterization of top-ranked candidates within the Co–Fe–Mn–Ni–Cu system represent a natural validation step.
The reported hypervolume plateau depends on the chosen reference point and normalization. More systematic sensitivity analyses are required to ensure robust conclusions about convergence, and additional tests exploring alternative reference definitions and scaling schemes are provided in the Supplementary Information. 

In summary, we have developed an ensemble-based multi-objective Bayesian optimization framework to accelerate the discovery of mechanically hard yet magnetically soft high-entropy alloys. By integrating a heterogeneous surrogate ensemble with Monte Carlo sampling, the framework achieves data-efficient convergence within approximately fifteen iterations, yielding a robust Pareto set of candidates that balance magnetic and mechanical properties. The optimization consistently directed the search toward a physically meaningful subspace, while systematically excluding less favorable elements such as Zn, Ti, and V. Further analysis reveals that descriptors play a decisive role in designing mechanically hard soft magnetic HEAs. 

\section{Computational methods}

% \subsection*{Pseudocode}
% Pseudocode

\begin{algorithm}[h!]
\caption{Meta-adaptive hybrid Bayesian optimization with Monte Carlo sampling}
\label{alg:meta-adaptive-hybrid-BO-converge}
\scriptsize
\begin{algorithmic}[1]
\Require 
    Training data $(X_{\mathrm{train}}, Y_{\mathrm{train}})$; candidate pool $X_{\mathrm{cand}}$;
    surrogate model set $\mathcal{M} = \{M_1, \ldots, M_k\}$; bootstrap size $B$; 
    maximum iterations $N$; acquisition function $\mathrm{ACQ}(\cdot)$;
    blending factor $\alpha \in [0,1]$
\Statex \textbf{GoalType:} $\mathrm{TargetVal}$ or $\mathrm{Max/Min}$; optionally specify \textbf{targetValue}, \textbf{targetThreshold}
\Ensure 
    Final updated dataset $(X_{\mathrm{train}}, Y_{\mathrm{train}})$

\State \textbf{Initialize:} noImprovementCount $\leftarrow 0$; bestVal $\leftarrow -\infty$ (or $+\infty$ for minimization)
\For{$t = 1 \to N$}
    \For{$M \in \mathcal{M}$}
        \State $\mathrm{bag}_M,\; \mathrm{r2}_M \leftarrow \emptyset,\; \emptyset$
        \For{$b = 1 \to B$}
            \State Sample bootstrap data for model $M$
            \State Train submodel $\mathrm{model}_{M,b}$; compute OOB score $R^2_{M,b}$
            \State $\mathrm{bag}_M \gets \mathrm{bag}_M \cup \{\mathrm{model}_{M,b}\}$; 
                   $\mathrm{r2}_M \gets \mathrm{r2}_M \cup \{R^2_{M,b}\}$
        \EndFor
    \EndFor

    \State \textbf{(A) OOB-based score:} 
    \State $\quad w'_M \gets \dfrac{1}{B} \sum_{b=1}^B R^2_{M,b}$

    \State \textbf{(B) Stacking-based score:} 
    Combine all $\mathrm{bag}_M$ into a single stacked RF; compute $w''_M$

    \State \textbf{(C) Final weight:} 
    \State $\quad \hat{w}_M = \alpha\,w'_M + (1-\alpha)\,w''_M$ (normalized over all $M$)

    \State \textbf{(D) Aggregated mean/variance:} 
    \State $\quad \bar{\mu}_M,\bar{\sigma}_M \gets \text{average over } B \text{ submodels}$  
    \State $\quad \mu(x) = \sum_{M} \hat{w}_M \bar{\mu}_M(x),\quad \sigma(x) = \sum_{M} \hat{w}_M \bar{\sigma}_M(x)$

    \State \textbf{(E) Acquisition \& Monte Carlo Update:}
    \State \quad Let $\mathrm{aggModel}(x) = \mu(x)$  \Comment{Aggregated predictive mean}
    \State \quad If maximizing, either pass $-\mathrm{aggModel}$ to Monte Carlo, or handle negation internally
    \State \quad $X_{\mathrm{MC}} = \mathrm{MONTE\_CARLO\_SAMPLING}(\mathrm{aggModel},\; X_{\mathrm{cand}},\; \mathrm{MC\_Params})$
    \State \quad $x_{\mathrm{new}} \gets X_{\mathrm{MC}}[0]$ \Comment{Select the best candidate}
    \State \quad $y_{\mathrm{new}} \gets \textsc{Evaluate}(x_{\mathrm{new}})$
    \State \quad $X_{\mathrm{train}} \gets X_{\mathrm{train}} \cup \{x_{\mathrm{new}}\}$
    \State \quad $Y_{\mathrm{train}} \gets Y_{\mathrm{train}} \cup \{y_{\mathrm{new}}\}$

    \State \textbf{(F) Convergence check:}
    \If{\textbf{GoalType} = $\mathrm{TargetVal}$ \textbf{and} $|y_{\mathrm{new}} - \mathrm{targetValue}| \leq \mathrm{targetThreshold}$}
        \State \textbf{break} \Comment{Target reached}
    \EndIf
    \If{\textbf{GoalType} = $\mathrm{Max/Min}$ \textbf{and} no improvement in 2 iterations \textbf{and} decreasing model variance}
        \State \textbf{break} \Comment{Converged}
    \EndIf

    \If{$y_{\mathrm{new}}$ improves over bestVal}
        \State Update bestVal
    \Else
        \State noImprovementCount++
    \EndIf
\EndFor

\Return $(X_{\mathrm{train}}, Y_{\mathrm{train}})$
\end{algorithmic}
\end{algorithm}

\subsection*{Bayesian optimization}

We start with an initial training set of 10 randomly selected compositions. In each iteration, the model proposes 10 new compositions that are predicted to meet the target criteria. If any of these candidates achieve the desired outcome after computational evaluation, the optimization process terminates. If not, the new data is added to the training set, and the BO cycle continues. To effectively capture the complex interactions among components, we use Magpie descriptors~\cite{wardGeneralpurposeMachineLearning2016}, which provide detailed chemical information and composition-dependent statistics. The pseudo-code is summarized in the algorithm~\ref{alg:meta-adaptive-hybrid-BO-converge}

\subsubsection*{Multi-objective optimization}

Many real-world optimization tasks involve multiple conflicting objectives that must be addressed simultaneously, known as multi-objective optimization problems (MOPs). Mathematically, a MOP can be formulated as:
% Requires: \usepackage{amsmath}
\begin{equation}
    % \begin{aligned}
        \min_{\mathbf{x}} \quad \mathbf{f}(\mathbf{x}) = (f_1(\mathbf{x}), f_2(\mathbf{x}), \ldots, f_m(\mathbf{x})),  \quad \mathbf{x} \in \mathcal{X}
    % \end{aligned}
    \label{eq:multi_objective}
\end{equation}

In this formulation, $\mathbf{x} = (x_1, x_2, \ldots, x_d)$ represents the decision vector with $d$ variables, $\mathcal{X}$ is the decision space, and $\mathbf{f}$ comprises $m$ objectives where $m \geq 2$. When the number of objectives exceeds three, the problem is referred to as a many-objective optimization problem. The objective is to identify a set of optimal solutions that balance the different objectives, known as Pareto optimal solutions. The collection of all Pareto optimal solutions in the decision space is called the Pareto set, and their projection in the objective space forms the PF. The goal of multi-objective optimization is to obtain a representative subset of the PF.

\subsubsection*{Surrogate model}

We developed a tailored ensemble surrogate model that combines bootstrapping and stacking strategies~\cite{ting1997StackingBagged} to improve robustness in sparse data regimes. This framework aggregates heterogeneous base learners, each trained on different data subsets, and applies adaptive model weighting to effectively capture complex, nonlinear patterns in the high-dimensional, descriptor-based feature space. The ensemble incorporates a broad range of algorithms, including neural networks, LightGBM boosted trees~\cite{ke2017LightGBMHighly}, CatBoost boosted trees~\cite{dorogush2018CatBoostGradient}, Random Forests, Extremely Randomized Trees, and k-Nearest Neighbors. The rationale for employing such a diverse set of models is twofold: first, the ensemble achieves strong predictive performance by leveraging the complementary strengths of different algorithms, thereby enhancing prediction robustness; second, each base learner serves not only as a contributor to the ensemble’s accuracy but also as a sampler that captures unique data characteristics. This ensemble design promotes sampling diversity and mitigates the risk of over-reliance on any single model’s prediction, thus facilitating a more comprehensive exploration of the search space.

At each iteration, the surrogate model suggests 10 candidate compositions that are most likely to satisfy the predefined target criteria. For each selected surrogate algorithm, the procedure involves performing $B$ bootstrap sampling rounds to generate an ensemble of submodels. Each submodel is evaluated using its corresponding Out-of-Bag (OOB) $R^{2}$ score to quantify predictive reliability. Subsequently, the final weight $\omega_M$ of each surrogate algorithm is computed by integrating the average OOB score with a stacking-based weighting mechanism. This fusion is governed by a tunable parameter $\alpha$, which balances the interpolation and extrapolation capabilities of individual algorithms. These computed weights are then used to construct a unified ensemble predictor that provides both the predictive mean $\mu(x)$ and the associated uncertainty $\sigma(x)$ for each candidate composition $x\in X_{\text{cand}}$.

\subsection*{DFT calculations}

The DFT calculations employed the generalized gradient approximation for the exchange–correlation functional in the Perdew–Burke–Ernzerhof formulation \cite{perdew1996GeneralizedGradienta}. The Kohn–Sham equations were solved using the EMTO method \cite{vitos2001TotalenergyMethod,vitos2007ComputationalQuantum} combined with the CPA to treat chemical disorder \cite{vitos2001AnisotropicLattice}. A scalar-relativistic approximation with a soft-core basis was applied, and total energies were evaluated using the full charge density technique \cite{vitos1997FullChargedensity} within an optimized overlapping muffin-tin potential. Brillouin zone integrations were performed on a $17 \times 17 \times 17$ k-point grid. Magnetic properties were evaluated in both FM and PM states. The PM state is simulated using the DLM approximation \cite{gyorffy1985FirstprinciplesTheory}. The Curie temperature $T_c$ was estimated using the mean-field approximation, which relies on ab initio energy differences combined with simplified models from previous studies~\cite{kormann2015TreasureMaps, sato2003CurieTemperatures}.

\medskip
\textbf{Conflicts of interest}

The authors declare no conflict of interest.

\medskip
\textbf{Supporting Information} \par %Please delete the Suppporting Information statement if it is not applicable. Please supply Supporting Information in another file. Supporting information should not be provided in .tex format
Supporting Information is available from the Wiley Online Library or from the author.

% Acknowledgements
\medskip
\textbf{Acknowledgements} \par %delete if not applicable))

We acknowledge the support of the Collaborative Research Centre/Transregio (CRC/TRR) 270 funded by the Deutsche Forschungsgemeinschaft (DFG) and the CoCoMag project funded by the European Innovation Council (EIC). We also appreciate the computing time provided on the high-performance computer Lichtenberg at the NHR Centers NHR4CES at TU Darmstadt. X. L. and S. S. acknowledge the support of the Swedish Research Council and the \AA Forsk Foundation, the Swedish Energy Agency, and the Carl Tryggers Stiftelse Foundation.

\medskip
\textbf{Data Availability Statement}
Data supporting the findings of this study are available at \url{https://github.com/Daimian/HEA}.
% References
\medskip

% \appendix
% \section{Example Appendix Section}


%%       \citet{<label>} ==> Lamport (1994)
%%       \citep{<label>} ==> (Lamport, 1994)
%%
% Example citation, See \citet{lamport94}.

\bibliographystyle{elsarticle-num} 
\bibliography{mian}




% %% For authoryear reference style

% \bibitem[Lamport(1994)]{lamport94}
%   \textit{\LaTeX: a document preparation system},
%   Addison Wesley, Massachusetts,

% \end{thebibliography}
\end{document}